\documentclass[a4paper,11pt]{article}
\usepackage{pos}
\usepackage{bm}
\usepackage{braket}
\usepackage{wrapfig}

\newcommand{\gtapprox}{\raisebox{-0.5ex}{$\,\stackrel{>}{\scriptstyle\sim}\,$}}

\title{Relativistic corrections to the static potential from generalized Wilson loops at finite flow time}
\ShortTitle{Relativistic corrections to the static potential at finite flow time}

\author*[a,b]{Michael Eichberg}
\author[a,b]{Marc Wagner}

\affiliation[a]{Institut f\"ur Theoretische Physik, Goethe University,\\
  Max-von-Laue-Straße 1, D-60438 Frankfurt am Main, Germany}

\affiliation[b]{Helmholtz Research Academy Hesse for FAIR, Campus Riedberg, \\
  Max-von-Laue-Straße 12, D-60438 Frankfurt am Main, Germany}

\emailAdd{eichberg@itp.uni-frankfurt.de}
\emailAdd{mwagner@itp.uni-frankfurt.de}

\abstract{We present results from an ongoing project concerned with the computation of $\mathcal{O}(1/m_Q)$ and $\mathcal{O}(1/m_Q^2)$ relativistic corrections to the static potential. These corrections are extracted from Wilson loops with two chromo-field insertions. We use gradient flow, which allows to renormalize the inserted fields and leads to a significantly improved signal-to-noise ratio, providing access to loops with large spatial and temporal extents.}

\FullConference{The 40th International Symposium on Lattice Field Theory (Lattice 2023)\\
July 31st - August 4th, 2023\\
Fermi National Accelerator Laboratory\\}

\begin{document}
\maketitle

\section{Introduction}

  A standard method to predict masses and properties of heavy quarkonia is to solve a Schr\"odinger equation with a suitable heavy quark-anti-quark potential. A simple choice would be the ordinary static potential, which is, however, a crude approximation for finite heavy quark mass $m_Q$ and e.g.\ incapable to describe spin-splitting. Such effects can be included by considering relativistic corrections to the static potential. In the past, relativistic corrections proportional to $1/m_Q$ and $1/m_Q^2$ have been derived in numerous ways, most prominently in potential Non-Relativistic QCD (pNRQCD), which is an effective field theory \cite{Pineda:2000sz}.
  
 The potential corrections from pNRQCD have been calculated perturbatively up to N$^3$LO and N$^3$LL accuracy \cite{Peset:2015vvi,Peset:2018jkf,Anzai:2018eua}. Non-perturbative lattice gauge theory computations are possible, but technically difficult (see Refs.\ \cite{deForcrand:1985zc,Campostrini:1986ki,Bali:1997am,Koma:2006fw}). One reason for that are strong UV fluctuations in the corresponding chromo-field correlators. Moreover, it is necessary to renormalize the chromo-field insertions. To overcome these difficulties, we use gradient flow \cite{Luscher:2010iy}, which is useful for renormalization and strongly suppresses UV noise.


\section{The heavy quark-anti-quark potential with $1 / m_Q$ and $1 / m_Q^2$ corrections}

\begin{figure}[htb]
  \centering
  \includegraphics[width=0.25\textwidth]{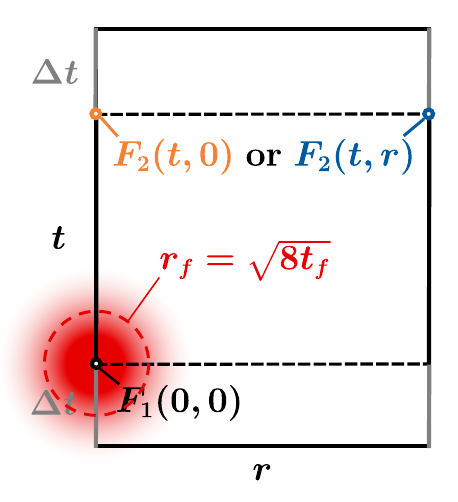}
  \caption{\label{fig:skizze}Generalized Wilson loop $W_{r \times (t + 2 \Delta t)}(F_2(t,r) F_1(0,0))$ with chromo-field insertions $F_1$ and $F_2$. The red circle has radius $r_f$ and marks the region, where the field $F_1$ is smeared by applying gradient flow with flow time $t_f = r_f^2 / 8$.}
\end{figure}

In pNRQCD the potential corresponding to a heavy quark-anti-quark pair, where both quarks have the same mass $m_Q$, can be written as
\begin{align}
V(r) = V^{(0)}(r) + \frac{1}{m_Q} V^{(1)}(r) + \frac{1}{m_Q^2} \left( V^{(2)}_\mathrm{SD}(r) + V^{(2)}_\mathrm{SI}(r) \right) + \mathcal{O}(1/m_Q^3)
\end{align}
\cite{Pineda:2000sz}. $V^{(0)}$ is the ordinary static potential, whereas $V^{(1)}$ and $V^{(2)}_\mathrm{SI}$ are spin-independent (SI) corrections and $V^{(2)}_\mathrm{SD}$ is a spin-dependent (SD) correction. $V^{(2)}_\mathrm{SI}$ and $V^{(2)}_\mathrm{SD}$ can be decomposed further with detailed equations presented in Ref.\ \cite{Pineda:2000sz}. The potential corrections can be calculated by solving integrals of the form
\begin{align}
\label{eq:corrintegral} \int_0^\infty dt \, t^s \langle \Sigma_g^+ , 0 | F_2(t,r) F_1(0,0) | \Sigma_g^+ , 0 \rangle , ~~~ s = 0,1,2 ,
\end{align}
\footnote{For $F_1 = F_2 = E_z$ (where we assume that the heavy quarks are separated along the $z$ axis) one has to subtract $\langle \Sigma_g^+ , 0 | E_z(t,r) | \Sigma_g^+ , 0 \rangle \langle \Sigma_g^+ , 0 | E_z(0,0) | \Sigma_g^+ , 0 \rangle$ from the expectation value.}
where $\langle \Sigma_g^+ , 0 | F_2(t,r) F_1(0,0) | \Sigma_g^+ , 0 \rangle$ denotes the $\Sigma_g^+$ ground state expectation value of two chromo-fields $F_1$ and $F_2$, i.e.\ the expectation value in the presence of a static quark-anti-quark pair and a flux tube corresponding to the ordinary static potential. $\langle \Sigma_g^+ , 0 | F_2(t,r) F_1(0,0) | \Sigma_g^+ , 0 \rangle$ can be expressed in terms of Wilson loop expectation values,
\begin{align}
\label{EQN001} \langle \Sigma_g^+ , 0 | F_2(t,r) F_1(0,0) | \Sigma_g^+ , 0 \rangle = \lim_{\Delta t \rightarrow \infty} \frac{W_{r \times (t + 2 \Delta t)}(F_2(t,r) F_1(0,0))}{W_{r \times (t + 2 \Delta t)}} ,
\end{align}
where $W_{r \times (t + 2 \Delta t)}(F_2(t,r) F_1(0,0))$ is a Wilson loop with two chromo-field insertions as shown in Fig.~\ref{fig:skizze} and $W_{r \times (t + 2 \Delta t)}$ is an ordinary Wilson loop of the same spatial and temporal extent. For the chromo-fields we use the common clover definition,
\begin{align}
	E_i = \frac{1}{2i} (\Pi_{i0} - \Pi^\dagger_{i0}), ~~~ B_i = \frac{\epsilon_{ijk}}{2i} (\Pi_{ij} - \Pi^\dagger_{ij}), ~~~ \Pi_{\mu\nu} = \frac{1}{4} (P_{\mu,\nu} + P_{\nu,-\mu} + P_{-\mu,-\nu} + P_{-\nu,\mu})
  \label{eq:clover}
\end{align}
with plaquettes $P_{\mu,\nu}$. The right hand side of Eq.\ (\ref{EQN001}) is then suited for lattice gauge theory computations.

\section{\label{SEC001}Gradient flow and renormalization}

In Refs.\ \cite{Bali:1997am,Koma:2006fw} chromo-fields were renormalized approximately, taking into account divergencies up to three loops, by multiplying each field insertion with a Huntley-Michael (HM) renormalization factor \cite{Huntley:1986de}. However, the statistically rather precise results for the potential corrections from Ref.~\cite{Koma:2006fw} did not fulfill the Gromes relation \cite{Gromes:1984ma,Brambilla:2004jw} indicating that HM renormalization is not appropriate for precision computations.

A modern method of renormalization is to use gradient flow \cite{Luscher:2010iy} and for chromo-magnetic fields also perturbatively calculated matching coefficients \cite{Pineda:2000sz}. Gradient flow generates a smooth gauge field $B_\mu$ and is defined via the flow equation
\begin{align}
  \dot{B}_\mu
  =
  D_\nu G_{\mu\nu} ,
\end{align}
where
\begin{eqnarray}
  B_\mu\Big|_{t_f=0} 
  =
  A_\mu , ~~~
  G_{\mu\nu} 
  =
  \partial_\mu B_\nu - \partial_\nu B_\mu + [B_\mu, B_\nu], ~~~
  D_\mu 
  =
  \partial_\mu + [B_\mu, \cdot] .
\end{eqnarray}
In particular for flow radius $r_f = \sqrt{8 t_f} \gtapprox a$ the $a$ dependence of the inserted chromo-fields is significantly reduced (see e.g.\ Refs.\ \cite{Leino:2021vop,Mayer-Steudte:2022uih}). Thus, a promising strategy to arrive at reliable and precise results for renormalized continuum extrapolated correlators $\langle \Sigma_g^+ , 0 | F_2(t,r) F_1(0,0) | \Sigma_g^+ , 0 \rangle$ is to carry out computations for several values of the flow time $t_f$ and the lattice spacing $a$. For chromo-magnetic field insertions the corresponding matching coefficients have to be included. Since gradient flow drastically reduces the $a$ dependence, a rather precise continuum extrapolation for each considered flow time $t_f$ is expected to be possible. These continuum extrapolated results can then be used for another extrapolation to flow time $t_f = 0$.

Gradient flow has also the benefit of suppressing UV fluctuations in chromo-field correlators leading to smaller statistical errors. However, gradient flow also comes with a caveat, namely that it corresponds to a smearing of the gauge field with approximate radius $r_f$ (see Fig.~\ref{fig:skizze}). If the flow time is chosen too large, the two inserted chromo-fields might overlap. This, in turn, causes sizable unwanted effects, such that the results are not anymore useful for the continuum and zero flow time extrapolation discussed above. To avoid such an ``oversmearing'', we exclusively consider generalized Wilson loops with spatial extent $r > 2 r_f$.


\section{Computation of the integral in Eq.\ (\ref{eq:corrintegral})}

  A strategy to solve the integral in Eq.\ (\ref{eq:corrintegral}) is to compute the correlator \\ $\langle \Sigma_g^+ , 0 | F_2(t,r) F_1(0,0) | \Sigma_g^+ , 0 \rangle$ at a number of discrete $t$ separations and fit an ansatz motivated by the spectral decomposition
\begin{align}
\label{eq:specdecomp} \langle \Sigma_g^+ , 0 | F_2(t,r) F_1(0,0) | \Sigma_g^+ , 0 \rangle = \sum_{\Lambda_\eta^\epsilon , n} \bra{\Sigma_g^+ , 0} \hat{F}_1 \ket{\Lambda_\eta^\epsilon , n} \bra{\Lambda_\eta^\epsilon , n} \hat{F}_2 \ket{\Sigma_g^+ , 0} e^{-\Delta E_{\Lambda_\eta^\epsilon , n} t}
\end{align}
\cite{Bali:1997am} with $\Delta E_{\Lambda_\eta^\epsilon , n} = E_{\Lambda_\eta^\epsilon , n} - E_{\Sigma_g^+ , 0}$. We note that symmetry arguments allow to restrict the sum over $\Lambda_\eta^\epsilon$ to only two sectors depending on the spatial separation axis and the inserted chromo-fields. We plan to discuss this in detail in an upcoming publication. Once the energy differences and amplitudes appearing on the right hand side of Eq.\ (\ref{eq:specdecomp}) are determined by the fit, the expression can be used in Eq.\ (\ref{eq:corrintegral}) and the integral can be solved analytically.


\section{Numerical results}

In the following we discuss first results for a single ensemble of 2000 gauge link configurations with $(T/a) \times (L/a)^3 = 60 \times 30^3$ lattice sites generated with the standard Wilson plaquette action using \texttt{CL2QCD} \cite{Bach:2021cl2qcd}. The gauge coupling $\beta = 6.451$ corresponds to lattice spacing $a = 0.048 \, \text{fm}$, when defining $r_0 = 0.5 \, \text{fm}$. To increase the ground state overlap, the spatial lines of all Wilson loops, without and with chromo-field insertions, were APE smeared with parameters $N_\text{APE} = 100$ and $\alpha_\text{APE} = 0.5$ (see Ref.\ \cite{Jansen:2008si} for detailed equations). For the error analysis \texttt{pyerrors} \cite{Joswig:2022qfe} with techniques from Refs.\ \cite{Wolff:2003sm,Ramos:2018vgu,Schaefer:2010hu} was used.

\begin{figure}
  \centering
  \includegraphics[width=0.48\textwidth]{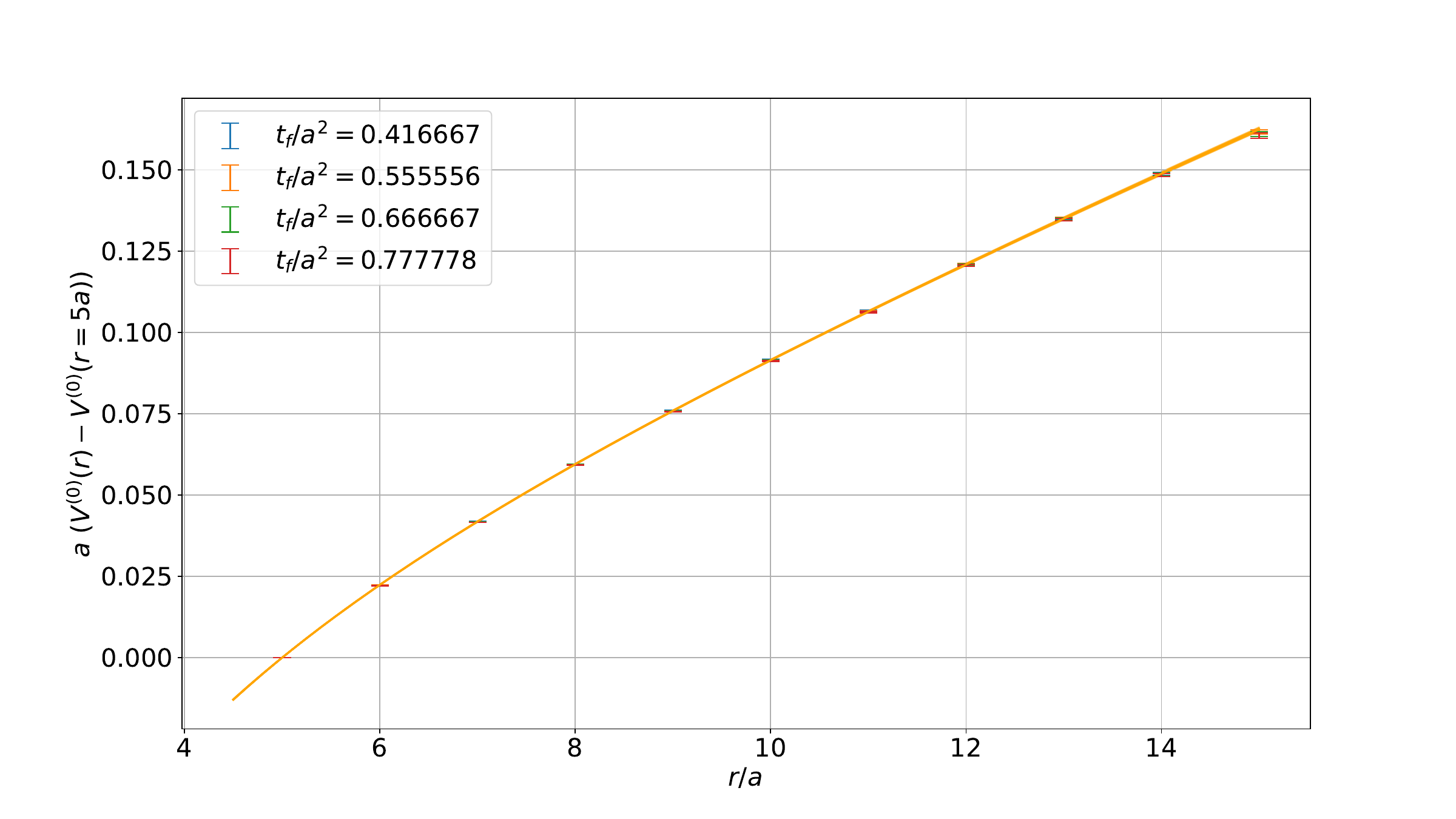}
  \caption{\label{fig:V0plot}The ordinary static potential for several values of the flow time. Data points are essentially on top of each other and, thus, hard to distinguish visually. The orange curve is a fit with a Cornell ansatz, $V_\text{Cornell}^{(0)}(r) = -c/r + \sigma r$.}
\end{figure}

In Fig.\ \ref{fig:V0plot} we show the ordinary static potential extracted from Wilson loops computed at four values of the flow time $t_f / a = 0.417 , \ldots , 0.778$, corresponding the flow radii $r_f/a = 1.83 , \ldots , 2.49$. To eliminate the self energy of the static quarks, which is reduced by gradient flow and depends on the flow time, we plot the difference $V^{(0)}(r) - V^{(0)}(r = 5a)$. The data points for all four values of the flow time are essentially on top of each other, indicating that gradient flow with flow radius $r_f/a \leq 2.49$ has negligible effect on the ordinary static potential for $r/a \geq 5$. This confirms our statement from section~\ref{SEC001} that considering only Wilson loops with spatial extents $r > 2 r_f$ ``oversmearing'' via gradient flow is not an issue. 

\begin{figure}
  \centering
  \includegraphics[width=0.9\textwidth]{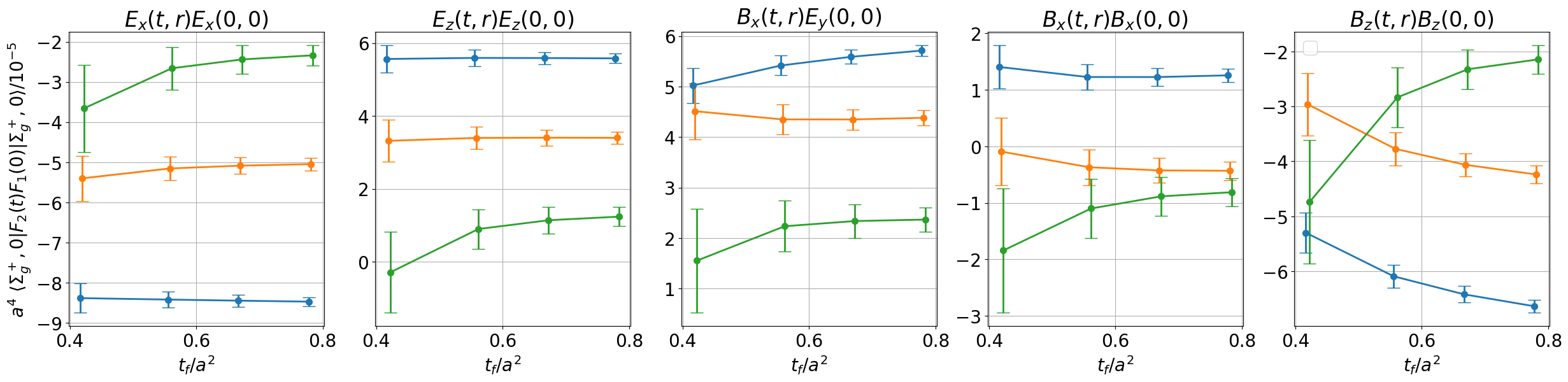}
  \caption{\label{fig:errorplot}Correlators $\langle \Sigma_g^+ , 0 | F_2(t,r) F_1(0,0) | \Sigma_g^+ , 0 \rangle$ as functions of the flow time $t_f$ for spatial separation $r/a = 8$ and different temporal separations $t/a = 5, 7, 10$.}
\end{figure}

In Fig.\ \ref{fig:errorplot} we present correlators $\langle \Sigma_g^+ , 0 | F_2(t,r) F_1(0,0) | \Sigma_g^+ , 0 \rangle$, which have been obtained by approximating the limit $\Delta t \rightarrow \infty$ on the right hand side of Eq.\ (\ref{EQN001}) by $\Delta t = 5 a$. The results are shown as functions of the flow time $t_f$. The three curves in each plot correspond to different temporal separations $t/a = 5, 7, 10$, whereas the spatial separation $r/a = 8$ is the same. Correlators with two chromo-electric field insertions appear to be independent of $t_f$ for $t_f/a \geq 0.417$ within statistical erors, while there might be a slight $t_f$ dependence for chromo-magnetic field insertions. The reason for the latter might be that we have not yet included the necessary matching coefficients for chromo-magnetic field insertions, which could weaken a possible $t_f$ dependence. The plots also demonstrate that gradient flow is able to drastically reduce statistical errors: increasing the flow time from $t_f / a = 0.417$ to $t_f / a = 0.778$, which corresponds to a rather mild increase of the flow radius from $r_f/a = 1.83$ to $r_f/a = 2.49$, reduces statistical errors by approximately a factor $5$.

\begin{figure}
  \centering
  \includegraphics[width=0.4\textwidth]{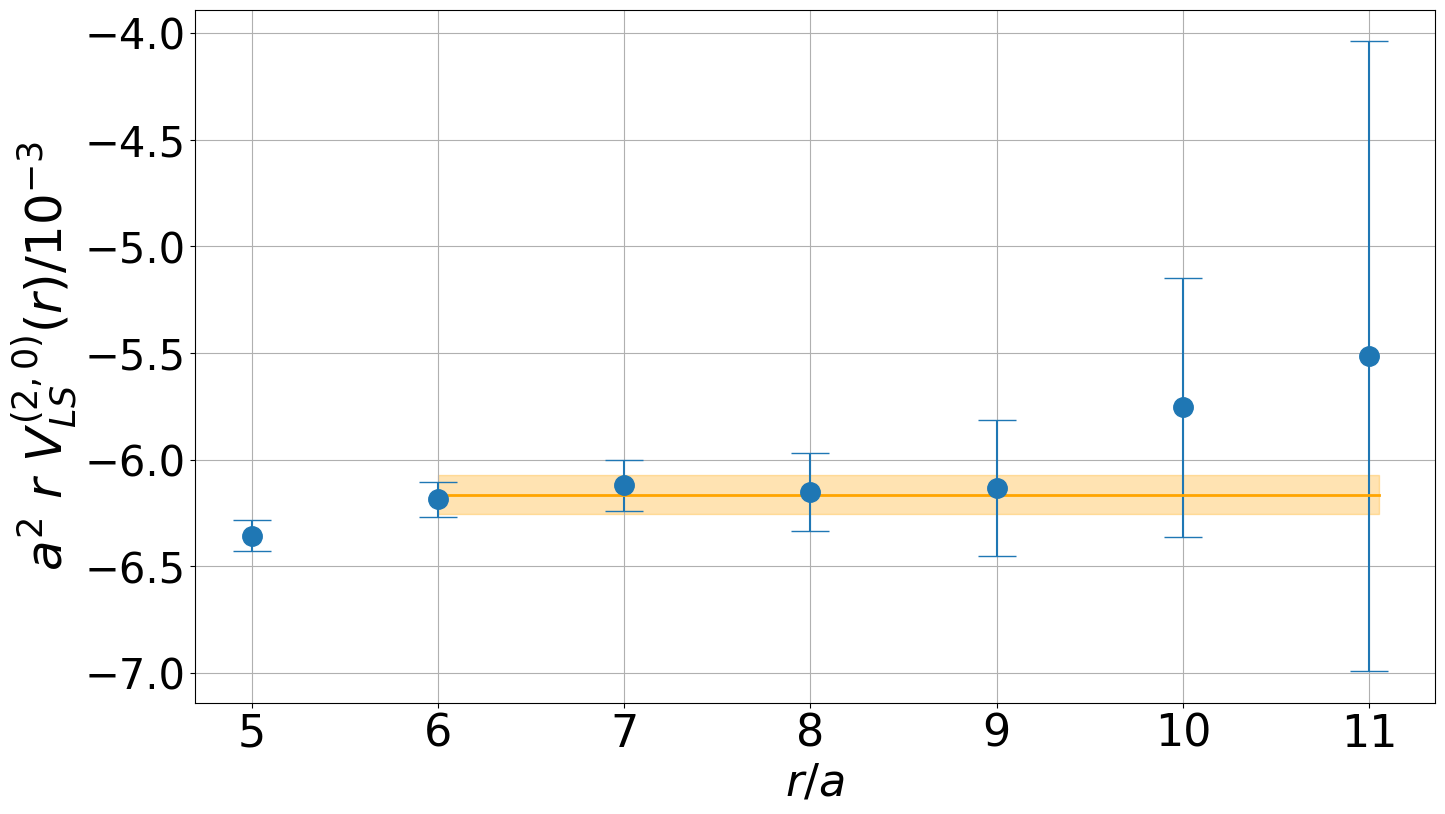}
  \includegraphics[width=0.4\textwidth]{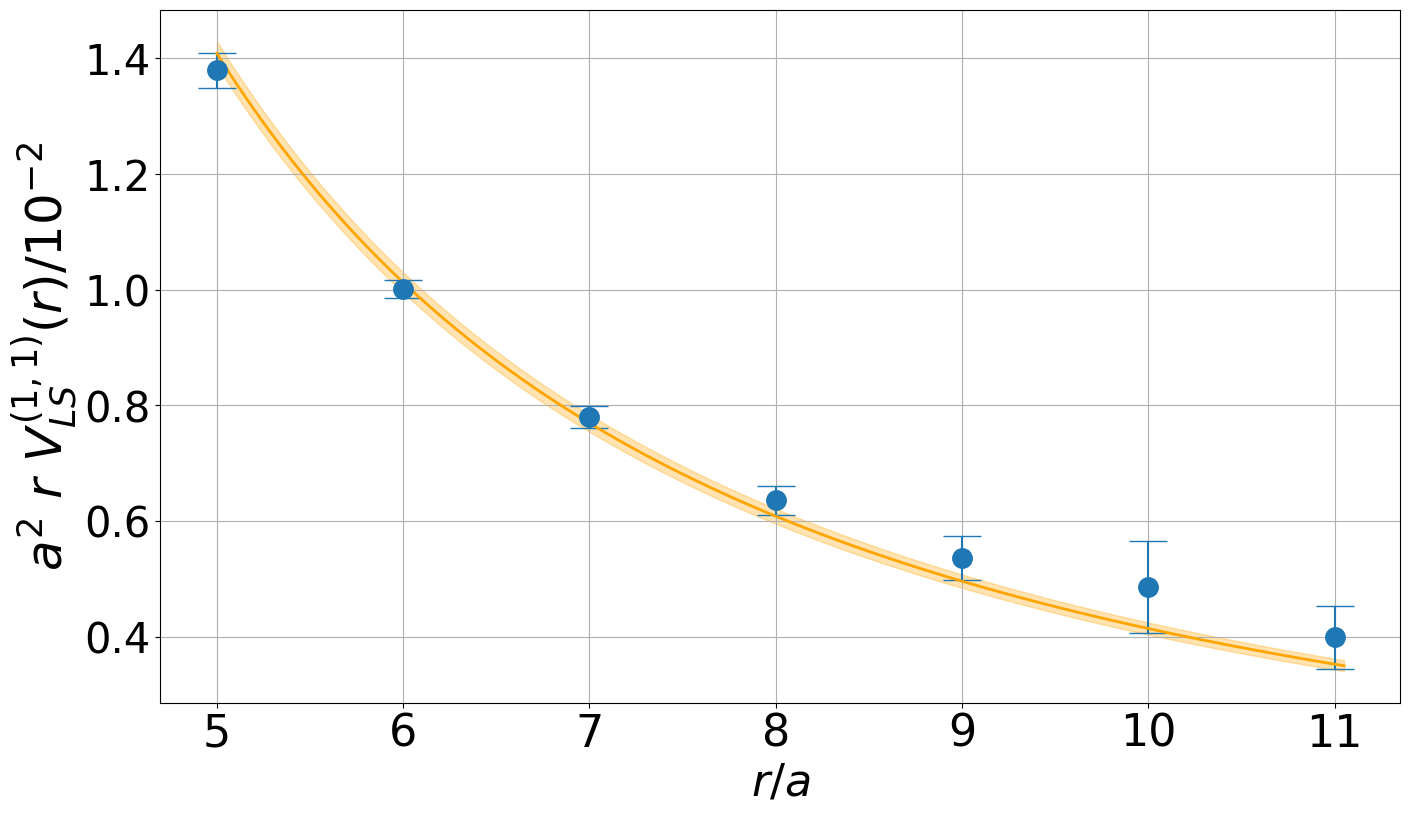}
  \includegraphics[width=0.4\textwidth]{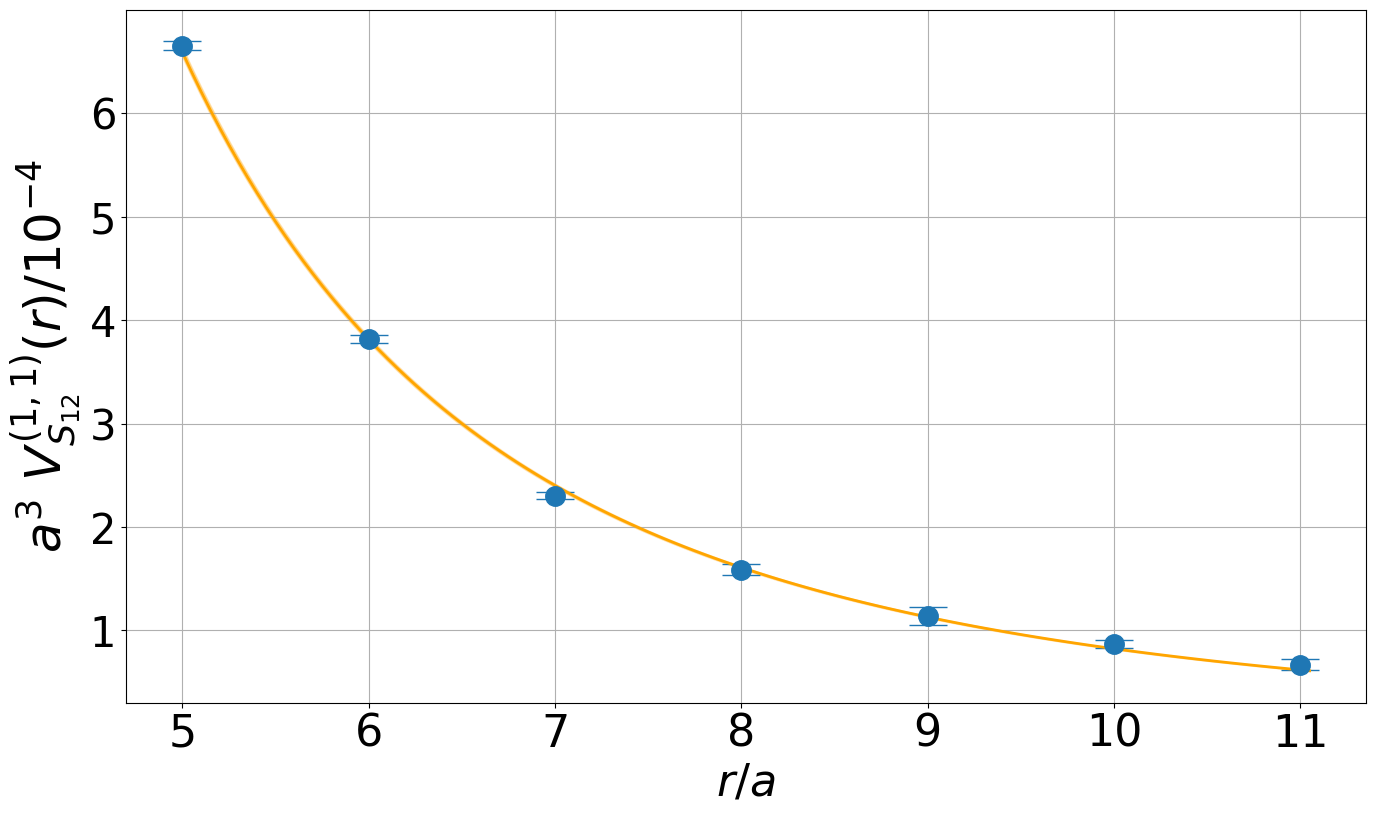}
  \includegraphics[width=0.4\textwidth]{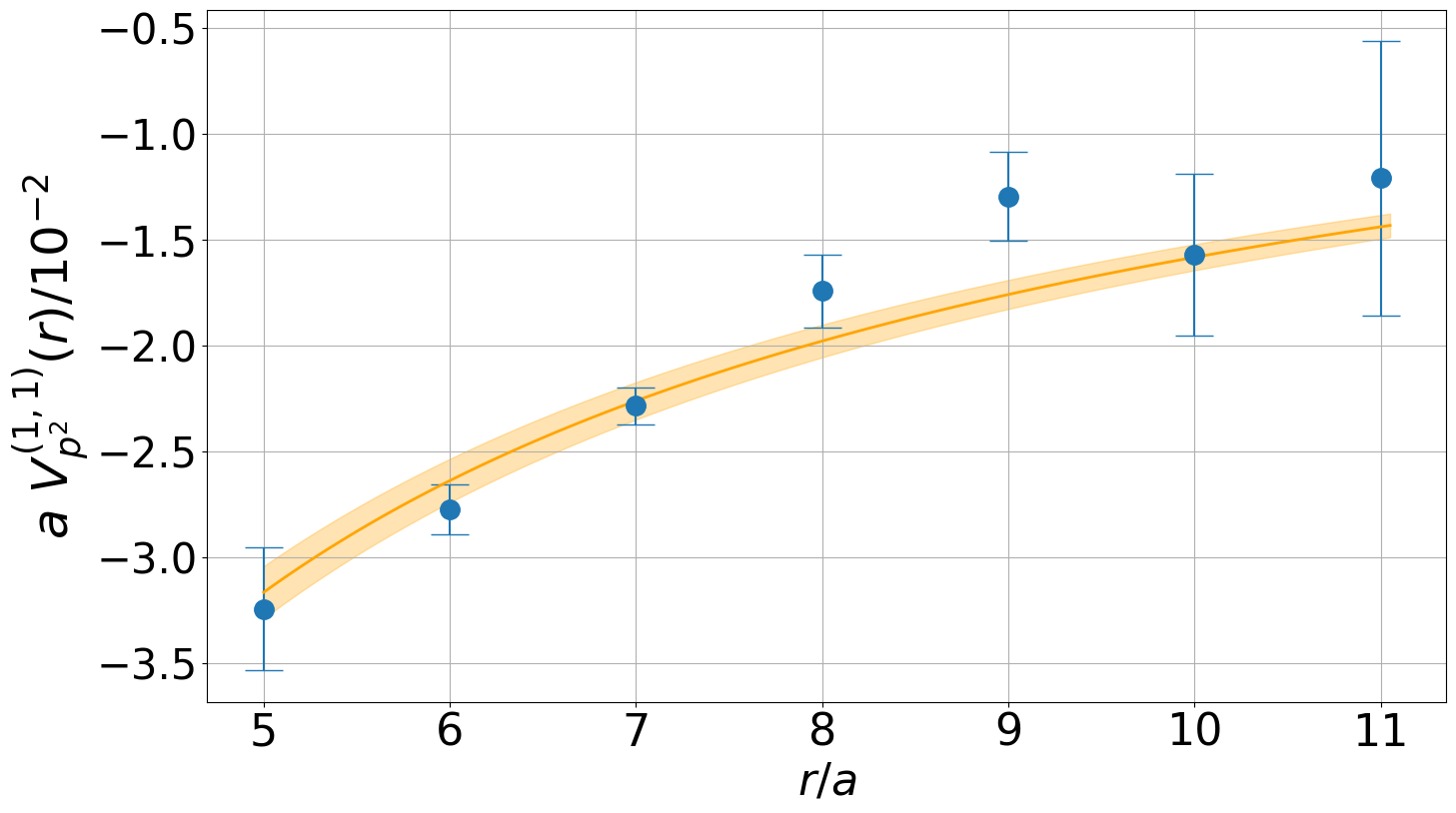}
  \caption{\label{fig:potentials}Selected potential corrections $V_{LS}^{(2,0)}$, $V_{LS}^{(1,1)}$, $V_{S_{12}}^{(1,1)}$ and $V_{p^2}^{(1,1)}(r)$ as functions of the spatial separation $r$ at flow time $t_f/a = 0.778$ and lattice spacing $a = 0.048 \, \text{fm}$.}
\end{figure}

We also generated crude results for potential corrections from correlators at finite flow time $t_f/a = 0.778$ and lattice spacing $a = 0.048 \, \text{fm}$. To this end, we again approximated the limit $\Delta t \rightarrow \infty$ on the right hand side of Eq.\ (\ref{EQN001}) by $\Delta t = 5 a$ and fitted the correlators with a truncated version of the spectral decomposition from Eq.\ (\ref{eq:specdecomp}). The integrals from Eq.\ (\ref{eq:corrintegral}) can then be solved analytically. Selected potential corrections $V_{LS}^{(2,0)}$, $V_{LS}^{(1,1)}$, $V_{S_{12}}^{(1,1)}$ and $V_{p^2}^{(1,1)}(r)$ are plotted in Fig.\ \ref{fig:potentials} as functions of the spatial separation $r$. Even though we have not yet included the necessary matching coefficients for chromo-magnetic field insertions, the potential corrections exhibit the expected behavior and can successfully be parameterized by the following functions
\begin{align}
  V_{LS}^{(2,0)}(r)~=~-\frac{\sigma}{2r}, ~~~
  V_{LS}^{(1,1)}(r)~=~\frac{c_F c}{r^3} - \frac{c_F g\Lambda^\prime}{r^2}, ~~~
  V_{S_{12}}^{(1,1)}(r)~=~\frac{c_F^2 c_{S_{12}}^2}{r^3}, ~~~
  V_{p^2}^{(1,1)}(r)~=~\frac{c_{p^2}}{2r},
\end{align}
which are similar to those used in Refs.\ \cite{Bali:1997am,Koma:2006fw} with an additional long range contribution for $V_{LS}^{(1,1)}$ as derived in Ref.\ \cite{Brambilla:2014eaa} ($c_F$ is the previously mentioned matching coefficient and $c$ and $\sigma$ are the parameters of the Cornell parameterization of the ordinary static potential $V^{(0)}$ [see caption of Fig.\ \ref{fig:V0plot}]). Moreover, statistical errors are quite small, indicating that we have a setup, which will allow a precise and rigorous determination of these potential corrections. We plan to carry out corresponding computations in the near future, which implies, in particular, to extrapolate the correlators to flow time $t_f = 0$ and to the continuum as discussed in section~\ref{SEC001}. Once these potential corrections are available, they could be used in Born-Oppenheimer predictions of the bottomonium and charmonium spectra. At a recent conference we have reported about a corresponding preparatory study \cite{Eichberg:2022zfv}.

\section*{Acknowledgements}

We acknowledge interesting and useful discussions with Nora Brambilla, Viljami Leino, Carolin Schlosser, Antonio Vairo, Xiang-Peng Wang.
M.W.\ acknowledges support by the Heisenberg Programme of the Deutsche Forschungsgemeinschaft (DFG, German Research Foundation) -- project number 399217702.
Calculations on the GOETHE-HLR and on the on the FUCHS-CSC high-performance computers of the Frankfurt University were conducted for this research. We thank HPC-Hessen, funded by the State Ministry of Higher Education, Research and the Arts, for programming advice.



\begin{thebibliography}{99}

\bibitem{Pineda:2000sz}
A.~Pineda and A.~Vairo,
Phys.\ Rev.\ D \textbf{63}, 054007 (2001)
[erratum: Phys.\ Rev.\ D \textbf{64}, 039902 (2001)]
[arXiv:hep-ph/0009145 [hep-ph]].

\bibitem{Peset:2015vvi}
C.~Peset, A.~Pineda and M.~Stahlhofen,
JHEP \textbf{05}, 017 (2016)
[arXiv:1511.08210 [hep-ph]].

\bibitem{Peset:2018jkf}
C.~Peset, A.~Pineda and J.~Segovia,
Phys.\ Rev.\ D \textbf{98}, no.\ 9, 094003 (2018)
[arXiv:1809.09124 [hep-ph]].

\bibitem{Anzai:2018eua}
C.~Anzai, D.~Moreno and A.~Pineda,
Phys.\ Rev.\ D \textbf{98}, 114034 (2018)
[arXiv:1810.11031 [hep-ph]].

\bibitem{deForcrand:1985zc}
P.~de Forcrand and J.~D.~Stack,
Phys.\ Rev.\ Lett.\ \textbf{55}, 1254 (1985)

\bibitem{Campostrini:1986ki}
M.~Campostrini, K.~Moriarty and C.~Rebbi,
Phys.\ Rev.\ Lett.\ \textbf{57}, 44 (1986)

\bibitem{Bali:1997am}
G.~S.~Bali, K.~Schilling and A.~Wachter,
Phys.\ Rev.\ D \textbf{56}, 2566-2589 (1997)
[arXiv:hep-lat/9703019 [hep-lat]].

\bibitem{Koma:2006fw}
Y.~Koma and M.~Koma,
Nucl.\ Phys.\ B \textbf{769}, 79-107 (2007)
[arXiv:hep-lat/0609078 [hep-lat]].

\bibitem{Luscher:2010iy}
M.~L\"uscher,
JHEP \textbf{08}, 071 (2010)
[erratum: JHEP \textbf{03}, 092 (2014)]
[arXiv:1006.4518 [hep-lat]].

\bibitem{Huntley:1986de}
A.~Huntley and C.~Michael,
Nucl.\ Phys.\ B \textbf{286}, 211-230 (1987).

\bibitem{Gromes:1984ma}
D.~Gromes,
Z.\ Phys.\ C \textbf{26}, 401 (1984).

\bibitem{Brambilla:2004jw}
N.~Brambilla, A.~Pineda, J.~Soto and A.~Vairo,
Rev.\ Mod.\ Phys.\ \textbf{77}, 1423 (2005)
[arXiv:hep-ph/0410047 [hep-ph]].

\bibitem{Leino:2021vop}
V.~Leino, N.~Brambilla, J.~Mayer-Steudte and A.~Vairo,
EPJ Web Conf.\ \textbf{258}, 04009 (2022)
[arXiv:2111.10212 [hep-lat]].

\bibitem{Mayer-Steudte:2022uih}
J.~Mayer-Steudte, N.~Brambilla, V.~Leino and A.~Vairo,
PoS \textbf{LATTICE2022}, 353 (2023)
[arXiv:2212.12400 [hep-lat]].

\bibitem{Bach:2021cl2qcd}
A.~Sciarra, C.~Pinke, M.~Bach, F.~Cuteri, L.~Zeidlewicz, C.~Sch\"afer, T.~Breitenfelder, C.~Czaban, S.~Lottini, P.F.~Depta,
CL2QCD (v1.1), Zenodo (2021),
https://doi.org/10.5281/zenodo.5121917.

\bibitem{Jansen:2008si}
K.~Jansen \textit{et al.} [ETM],
JHEP \textbf{12}, 058 (2008)
[arXiv:0810.1843 [hep-lat]].

\bibitem{Joswig:2022qfe}
F.~Joswig, S.~Kuberski, J.~T.~Kuhlmann and J.~Neuendorf,
Comput.\ Phys.\ Commun.\ \textbf{288}, 108750 (2023)
[arXiv:2209.14371 [hep-lat]].

\bibitem{Wolff:2003sm}
U.~Wolff [ALPHA],
Comput.\ Phys.\ Commun.\ \textbf{156}, 143-153 (2004)
[erratum: Comput.\ Phys.\ Commun.\ \textbf{176}, 383 (2007)]
[arXiv:hep-lat/0306017 [hep-lat]].

\bibitem{Ramos:2018vgu}
A.~Ramos,
Comput.\ Phys.\ Commun.\ \textbf{238}, 19-35 (2019)
[arXiv:1809.01289 [hep-lat]].

\bibitem{Schaefer:2010hu}
S.~Schaefer \textit{et al.} [ALPHA],
Nucl.\ Phys.\ B \textbf{845}, 93-119 (2011)
[arXiv:1009.5228 [hep-lat]].

\bibitem{Brambilla:2014eaa}
N.~Brambilla, M.~Groher, H.~E.~Martinez and A.~Vairo,
Phys.\ Rev.\ D \textbf{90}, 114032 (2014)
[arXiv:1407.7761 [hep-ph]].

\bibitem{Eichberg:2022zfv}
M.~Eichberg and M.~Wagner,
PoS \textbf{FAIRness2022}, 014 (2023)
[arXiv:2208.09337 [hep-lat]].
\end{thebibliography}
\end{document}